\documentclass[conference]{IEEEtran}
\IEEEoverridecommandlockouts
\usepackage{cite}
\usepackage{amsmath,amssymb,amsfonts}
\usepackage{algorithmic}
\usepackage{graphicx}
\usepackage{textcomp}
\usepackage[table]{xcolor}
\usepackage{xcolor}
\usepackage{url}
\usepackage{tabularx}
\usepackage{booktabs}
\def\BibTeX{{\rm B\kern-.05em{\sc i\kern-.025em b}\kern-.08em
    T\kern-.1667em\lower.7ex\hbox{E}\kern-.125emX}}
    
\begin{document}

\title{S2Accompanist: A Semantic-Aware and Structure-Guided Diffusion Model for Music Accompaniment Generation}


\author{
  \IEEEauthorblockN{Huakang Chen$^{1,*}$, Wenkai Cheng$^{1,*}$, Guobin Ma$^{1}$, 
  Chunbo Hao$^{1}$, Yuxuan Xia$^{1}$\\
  Mengqi Wei $^{1}$, Zhixian Zhao$^{1}$, Pengcheng Zhu$^{2}$, Hanbing Zhang$^{2}$, 
  Lei Xie$^{1,\dagger}$\thanks{$^{*}$Equal
  contribution. $^{\dagger}$Corresponding author.}}
  \IEEEauthorblockA{$^{1}$Audio, Speech and Language Processing Group
    (ASLP@NPU),\\
    School of Software, Northwestern Polytechnical University\\
    $^{2}$WeNet Open Source Community \\
  huakang@mail.nwpu.edu.cn, chengwenkai@mail.nwpu.edu.cn, lxie@nwpu.edu.cn}
}

\IEEEaftertitletext{\vspace{-1.5em}}

\maketitle

\begin{abstract}
High-fidelity text-to-music generation typically relies on massive proprietary datasets and immense computational resources. Existing models often struggle to generate coherent pure musical accompaniments and lack precise, localized semantic control due to their reliance on coarse, track-level annotations. To address these limitations under constrained data and computing resources, we propose S2Accompanist, a Semantic-Aware and Structure-Guided Diffusion Model developed for the ICME2026 ATTM Grand Challenge. Specifically, we design an automated data pipeline comprising structural segmentation, Large Audio-Language Model driven segment-level captioning, and dual-metric quality grading to overcome the absence of localized metadata in raw datasets. 
Furthermore, we propose a semantic-aware Variational Autoencoder fine-tuning strategy that explicitly distills foundational LeadSheet structures into the acoustic latent space, effectively improving the overall audio fidelity. Extensive experiments demonstrate that S2Accompanist achieves state-of-the-art objective performance on the ATTM Grand Challenge benchmark across both the Efficiency and Performance Tracks. With only 402M parameters, our model remains competitive compared to larger-scale unconstrained models and secured first place in the Efficiency Track.
\end{abstract}

\begin{IEEEkeywords}
Text-to-Music, Diffusion Model, Music Generation
\end{IEEEkeywords}

\section{Introduction}
\label{sec:intro}

The field of Text-to-Music (T2M) generation has experienced rapid advancements in recent years, largely driven by the scalability of diffusion models \cite{diffrhythmplus,ning2025diffrhythm,gong2025acestep,gong2026acestep15,noise2music,mousai,musicflow,musicldm} and autoregressive transformers \cite{musicgen,musiclm,yue,inspiremusic}. Leading open-source and proprietary models, such as Acestep series \cite{gong2025acestep,gong2026acestep15}, DiffRhythm series \cite{ning2025diffrhythm,diffrhythmplus}, and YuE \cite{yue}, have demonstrated remarkable capabilities in synthesizing high-quality music from natural language prompts. However, the success of these state-of-the-art systems is predominantly built upon massive, unconstrained proprietary datasets comprising tens of thousands of hours of music tracks \cite{gong2026acestep15}, alongside immense computational resources. 

Despite their impressive global performance, current T2M models exhibit notable limitations when applied to precise accompaniment generation. First, these models are primarily optimized for full mixed tracks containing both vocals and instruments. Consequently, they often struggle to synthesize purely instrumental music, frequently exhibiting vocal artifacts or structural collapse. Second, they typically rely on global, track-level descriptions that lack fine-grained structural alignments. The absence of localized metadata makes it exceedingly difficult for the models to accurately control local musical transitions, specific instrumentation, and emotional dynamics throughout different sections of a generated track.

To address the bottleneck of resource dependency and the critical need for controllable accompaniment generation, the ICME2026 ATTM Grand Challenge \cite{hsieh2026attm} establishes a rigorous benchmark. The challenge restricts model training exclusively to the moderately sized MTG-Jamendo dataset \cite{bogdanov2019mtg}, forcing a paradigm shift from simple data and compute scaling to intelligent architectural and data-engineering efficiency. Generating high-fidelity music under such limited data resources poses a significant challenge, particularly because the provided raw dataset severely lacks the fine-grained semantic descriptions and localized structural metadata necessary for precise condition modeling.

\begin{figure*}[htbp]
\centerline{\includegraphics{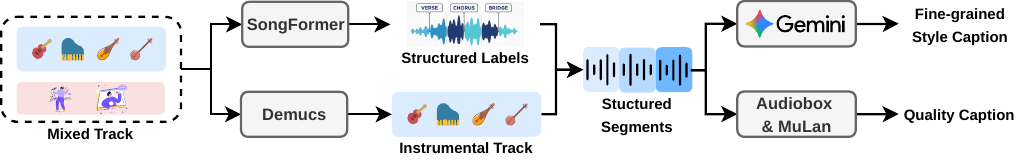}}
\caption{The proposed data pipeline. It extracts instrumental tracks, aligns them with structural timestamps derived from mixed audio, and generates fine-grained LALM captions followed by a dual-metric quality grading process.}
\label{fig:data}
\vspace{-12pt}
\end{figure*}

To bridge these gaps, we propose S2Accompanist, a Semantic-Aware and Structure-Guided Diffusion Model tailored for resource-efficient music accompaniment generation. Instead of training on randomly cropped audio snippets, we construct a comprehensive automated data pipeline that extracts localized, structurally discrete pure-music segments. We further enrich these segments with detailed semantic captions using a Chain-of-Thought (CoT) prompting strategy driven by a Large Audio-Language Model (LALM). To optimize the training data, we incorporate a dual-metric grading mechanism to evaluate these segments, ultimately yielding a top-scored subset specifically for the Supervised Fine-Tuning (SFT) phase. Additionally, to overcome the performance ceiling imposed by standard autoencoders, we introduce a semantic-aware Variational Autoencoder (VAE) fine-tuning technique that distills foundational LeadSheet representations into the acoustic latent space, which contributes to improved intrinsic musicality and harmonic consistency.

Our main contributions are summarized as follows:
\begin{itemize}
    \item \textbf{Structure-Guided Data Pipeline.} We construct an automated data pipeline based on structural segmentation, which extracts discrete music segments with fine-grained captions and performs quality grading to construct a refined subset for supervised fine-tuning.
    \item \textbf{Semantic-Aware VAE Fine-Tuning.} We develop a fine-tuning mechanism that distills LeadSheet-based structural semantics into the acoustic latent space. By explicitly embedding these musical skeletons, our approach consistently improves the fidelity and structural coherence of the generated audio.
    \item \textbf{S2Accompanist.} We present an efficient Diffusion Transformer (DiT) that fully exploits our proposed data pipeline. Trained under strictly limited resources, S2Accompanist generates high-fidelity accompaniments, outperforming larger unconstrained models and achieving the top position in the challenge's Efficiency Track.
\end{itemize}

\section{Methodology}
\label{sec:method}

\subsection{Data Pipeline}
\label{ssec:data}

To strictly comply with the ATTM Grand Challenge's fair-play constraints, our model is trained exclusively on the provided MTG-Jamendo dataset. However, raw instrumental tracks often lack the fine-grained semantic descriptions and structural metadata necessary for high-fidelity accompaniment generation. To address this, we design a comprehensive automated data pipeline, as illustrated in Figure \ref{fig:data}, comprising three main stages: instrumental extraction, structure labeling, and semantic captioning with quality-based data grading.

\textbf{Instrumental Extraction.} Following the challenge guidelines to focus on instrumental music generation, we first process the raw mixed tracks to eliminate vocal components. We utilize Demucs \cite{demucs} as an auxiliary source separation model to isolate the instrumental track from the original mixed track segments. This ensures that our generative model learns purely acoustic accompaniment features without being misled by residual vocal artifacts.

\textbf{Structure Labeling.} The composition of music accompaniments is inherently complex, exhibiting significant variations in instrumentation, emotional dynamics, and musical topics across different sections of a single track. Relying exclusively on global track-level annotations, which only reflect the dominant elements of a song, tends to mislead the generative model and severely limits its ability to learn precise and localized style control. To achieve fine-grained condition modeling, structural segmentation is essential. However, existing music structure analysis models are predominantly designed for full mixed songs containing vocals, lacking the capability to accurately segment pure accompaniment tracks. Fortunately, since the vast majority of the source MTG-Jamendo dataset originally consists of vocal-mixed tracks, we devise an elegant workaround. We first apply a structural segmentation model \cite{hao2025songformer} to the original mixed tracks to predict discrete structural tags (e.g., verse, chorus, bridge) and their corresponding accurate timestamps. Subsequently, we use these synchronized timestamps to slice the separated instrumental tracks. This strategic alignment ultimately yields high-quality, structurally discrete pure-music segments.

\textbf{Fine-Grained Semantic Captioning and Quality-Based Data Grading.} To obtain precise style descriptions, we design a CoT prompting strategy for Gemini 2.5 Pro \cite{gemini}, acting as our LALM. Rather than generating captions directly, we first instruct the LALM to identify and extract granular tag-level attributes across six predefined dimensions: genre, mood, instrument, scene, region, and topic. Subsequently, it synthesizes these structured tags into a comprehensive, natural-language style description. 
This two-step elicitation process is designed to improve the semantic density and accuracy of the resulting localized captions.
Furthermore, recognizing the inevitable audio quality degradation introduced during the instrumental extraction process (e.g., separation artifacts), alongside potential semantic inconsistencies or hallucinations in the LALM-generated annotations, we implement a dual-metric data grading mechanism. We evaluate every segment using Audiobox \cite{audiobox} to quantify objective audio quality and MuLan \cite{muq} to measure the text-audio semantic similarity. Based on these two dimensions, the dataset is stratified: the entire corpus of structurally annotated and captioned segments is leveraged during the initial pre-training phase to build a robust generative prior. For the subsequent SFT phase, we apply strict empirical thresholds on both scores, retaining only the top 20\% of the highest-quality data. This strategy ensures optimal text-to-music alignment and superior audio fidelity during fine-tuning.

\begin{figure*}[htbp]
\centerline{\includegraphics{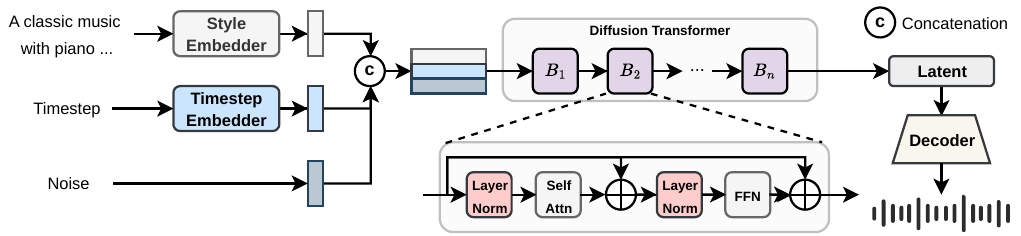}}
\caption{The overall model architecture of S2Accompanist. It utilizes a conditional DiT equipped with text/audio MuLan embeddings, optimized specifically on structurally discrete segments.}
\label{fig:dit}
\vspace{-12pt}
\end{figure*}

\subsection{Semantic-Aware VAE Fine-Tuning}
\label{ssec:vae}

The generation ceiling of a latent diffusion model is fundamentally constrained by the quality of its underlying autoencoder. In the context of pure accompaniment generation, we observe that existing VAEs exhibit notable limitations. For instance, the Stable Audio VAE \cite{stable_audio_vae} (adopted by systems like DiffRhythm \cite{ning2025diffrhythm} and DiffRhythm+ \cite{diffrhythmplus}) often struggles to preserve musicality, whereas the DiffRhythm2 VAE \cite{jiang2025diffrhythm2} employs a low frame rate that significantly compromises high-frequency reconstruction quality.

To overcome these bottlenecks, we propose a semantic-aware fine-tuning strategy tailored specifically for accompaniments. Our approach draws inspiration from two aspects: we utilize the Semantic VAE framework \cite{niu2025semanticvae} to explicitly inject fundamental structures into the acoustic latent space, and we draw inspiration from the \textit{Music-Semantic-VAE} project\footnote{\url{https://github.com/ASLP-lab/Music-Semantic-VAE}} by selecting the LeadSheet as our semantic target. As the foundational skeleton of a musical piece, the LeadSheet encapsulates complete modality, chord progressions, and rhythmic frameworks—elements that are indispensable for generating harmonious and structured accompaniments.

To implement this, we replace the pretrained Self-Supervised Learning (SSL) model used in the original Semantic VAE with SheetStage \cite{donahue2022sheetstage}, utilizing it as a frozen semantic teacher to extract robust LeadSheet representations from the input audio. The acoustic latent variables encoded by the VAE are directly projected through a Multi-Layer Perceptron (MLP) and aligned with the frozen SheetStage features using the Semantic Regularization Loss designed in the Semantic VAE \cite{niu2025semanticvae}. This distillation process forces the latent space to encode not only acoustic textures but also deep musical semantics.

In practice, to ensure high-fidelity audio quality, we first reproduce the DiffRhythm VAE architecture operating at a high frame rate of 25Hz as our base model. Subsequently, we perform semantic distillation on this base model using our curated MTG-Jamendo pure-music dataset. The fine-tuning objective jointly optimizes the Semantic Loss alongside standard VAE constraints, including reconstruction loss, Kullback-Leibler (KL) divergence, and adversarial discriminator loss. This results in a latent space that  preserves both acoustic fidelity and musical structure, laying a solid foundation for the downstream diffusion process.

\subsection{S2Accompanist DiT}
\label{ssec:model}

S2Accompanist employs a conditional DiT architecture adapted from DiffRhythm+, streamlined specifically for pure accompaniment generation by removing lyrics conditioning. As illustrated in Figure \ref{fig:dit}, during the forward process, we utilize MuLan as a style embedder to process our fine-grained semantic captions. The resulting style embedding, timestep embedding, and noisy latent are directly concatenated along the channel dimension and fed into the DiT blocks to predict the target latent, which is subsequently reconstructed into audio via our Semantic-Aware VAE. 

To fully leverage our customized data pipeline, the model is trained explicitly on the discrete structured segments (derived in Section \ref{ssec:data}) rather than conventional random audio crops. This enables the learning of localized and coherent musical progression. Furthermore, to accelerate convergence and mitigate cross-modal alignment difficulties, we employ a mixed-modality conditioning strategy during training by randomly alternating between audio-derived and text-derived MuLan embeddings at a 50\% ratio.

\begin{table*}[htbp]
\centering
\caption{Objective evaluation results on the final ATTM test prompts. S2Accompanist (Ours) corresponds to Submission e07 in the official leaderboard, achieving the best FAD overall and securing Rank 1 in the Efficiency Track.}
\label{tab:main_results}
\resizebox{\textwidth}{!}{
\begin{tabular}{lccccccc}
\toprule
\textbf{Model} & \textbf{Params} & \textbf{Track} & \textbf{Train Data} & \textbf{FAD $\downarrow$} & \textbf{CLAP $\uparrow$} & \textbf{CCS $\uparrow$} & \textbf{Rank} \\
\midrule
\multicolumn{8}{c}{\cellcolor[HTML]{E6F2FF}\textit{Pre-trained Reference Models (Unconstrained Proprietary Data)}} \\
\midrule
Stable Audio Open \cite{stable_audio_open} & 1.1B & - & 7.3K hrs & 0.574 & 0.321 & 0.800 & - \\
MusicGen-small \cite{musicgen} & 300M & - & 20K hrs & 0.574 & \textbf{0.370} & 0.875 & - \\
MusicGen-medium \cite{musicgen} & 1.5B & - & 20K hrs & 0.548 & 0.353 & \textbf{0.892} & - \\
\midrule
\multicolumn{8}{c}{\cellcolor[HTML]{E6F2FF}\textit{ATTM Challenge Submissions (Constrained Strictly to MTG-Jamendo Dataset)}} \\
\midrule
FluxAudio-S (Challenge Baseline) & 120M & Efficiency & 3.7K hrs & 0.757 & 0.088 & 0.592 & 17 \\
Submission p05 (Top Performance) & 2.4B & Performance & 0.46K hrs & 0.514 & \textbf{0.306} & 0.800 & 5 \\
Submission e01 & 189M & Efficiency & 3.7K hrs & 0.577 & 0.338 & 0.863 & 2 \\
Submission e05 & 499M & Efficiency & 0.46K hrs & 0.487 & 0.305 & 0.800 & 2 \\
Submission e08 & 450M & Efficiency & 3.7K hrs & 0.495 & 0.295 & 0.804 & 2 \\
\midrule
\textbf{S2Accompanist (Ours)} & \textbf{402M} & \textbf{Efficiency} & \textbf{3.7K hrs} & \textbf{0.417} & 0.261 & \textbf{0.867} & \textbf{1} \\
\bottomrule
\end{tabular}}
\vspace{-12pt}
\end{table*}

\section{Experiments}
\label{sec:exp}

\subsection{Training Details}
\label{ssec:training_details}

All experiments are conducted using two NVIDIA RTX A6000 GPUs. The training pipeline is divided into the optimization of the Semantic-Aware VAE and the S2Accompanist DiT. 

\textbf{Semantic-Aware VAE.} The architecture of our Semantic-Aware VAE is identical to that of DiffRhythm, which inherits from the Stable Audio 2 VAE. It comprises purely convolutional encoder and decoder blocks, totaling approximately 157M parameters. The model processes audio at a sampling rate of 24 kHz and applies downsampling factors of $[4, 5, 6, 8]$, yielding a compressed 64-dimensional latent representation at a frame rate of 25 Hz. For semantic fine-tuning, we slice the full MTG-Jamendo dataset into 3-second segments and optimize the model for 100k steps in full precision (FP32). Unmentioned hyperparameters strictly follow the configurations detailed in the original Semantic VAE.

\textbf{S2Accompanist DiT.} Our diffusion model is scaled to approximately 400M parameters with a hidden dimension of 1536 and 12 attention heads. The training consists of two phases: a pre-training phase utilizing the complete MTG-Jamendo dataset for 400k steps, followed by an SFT phase over 10 epochs, utilizing only the top 20\% high-quality data filtered by our dual-metric grading pipeline. During training, rather than using random audio crops, we explicitly sample the structured segments derived from our data pipeline. Specifically, we select structured segments with durations between 10 and 30 seconds; segments shorter than 10 seconds are discarded, while those exceeding 30 seconds are truncated to the 30-second maximum. To improve computational efficiency, the DiT is trained using FP16 half-precision, with optimization hyperparameters adopting the default settings of DiffRhythm.

\subsection{Evaluation Tasks and Metrics}
\label{ssec:eval_metrics}

To strictly adhere to the evaluation protocol of the ATTM Grand Challenge, we benchmark S2Accompanist against the official challenge baseline (FluxAudio-S) alongside representative state-of-the-art T2M models (e.g., MusicGen, Stable Audio Open \cite{stable_audio_open}). The objective evaluation is conducted on the official curated test prompts using three primary metrics:

\noindent\textbf{Fréchet Audio Distance (FAD)}\quad Measures the audio fidelity and distributional similarity between the generated samples and a hidden instrumental reference set from MTG-Jamendo. Following the challenge specifications, we utilize the \texttt{music\_audioset\_epoch\_15\_esc\_90.14} checkpoint of the CLAP-Laion-Music model \cite{clap} as the feature extractor. A lower FAD indicates superior acoustic quality.

\noindent\textbf{CLAP Score}\quad Evaluates the global semantic alignment between the input text prompts and the generated audio. We compute the cosine similarity in the joint embedding space using the identical CLAP checkpoint utilized for FAD calculation. A higher score reflects better overall text-to-music relevance.

\noindent\textbf{Concept Coverage Score (CCS)}\quad A novel, fine-grained semantic metric introduced by the challenge. It employs the Qwen3-Omni \cite{xu2025qwen3omni} LALM as a zero-shot judge to verify the presence of specific musical concepts (genre, instrument, and mood) within the generated audio based on output log-probabilities. Given that our S2Accompanist explicitly leverages a fine-grained semantic pipeline, CCS serves as a crucial indicator of our model's precise condition adherence. A higher CCS indicates superior concept-level semantic alignment.

\subsection{Objective Evaluation Results}
\label{ssec:obj_eval_results}

We present the objective evaluation results of S2Accompanist (officially submitted as \textit{Submission e07}) against both the challenge baselines and state-of-the-art pre-trained models. To ensure a comprehensive comparison, Table \ref{tab:main_results} highlights our model alongside the official challenge baseline, the top-performing submissions from both the Efficiency and Performance tracks, and established unconstrained models.

\textbf{Competitive Audio Fidelity:} 
As demonstrated in Table \ref{tab:main_results}, S2Accompanist achieves a FAD of 0.417, establishing a new state-of-the-art for this benchmark. Notably, our model not only outperforms all competing submissions within the 500M parameter limit (Efficiency Track) but also substantially surpasses large-scale models in the Performance Track (e.g., Submission p05 with 2.4B parameters). In addition, S2Accompanist eclipses unconstrained pre-trained systems such as MusicGen-medium and Stable Audio Open, which were trained on datasets up to five times larger. We attribute this high fidelity to two core methodological innovations: (1) our stringent dual-metric data grading pipeline, which isolates the top 20\% cleanest instrumental tracks for fine-tuning, thereby effectively eliminating acoustic separation artifacts; and (2) the Semantic-Aware VAE fine-tuning. 
By explicitly distilling foundational LeadSheet structures into the acoustic latent space, the VAE inherently preserves vital musical skeletons, which contributes to the overall audio quality and coherence of the reconstructed audio.

\textbf{Superior Semantic Alignment:}
In terms of semantic control, S2Accompanist achieves a CCS of 0.867, securing the highest fine-grained alignment score among all challenge participants. The CCS metric evaluates the zero-shot presence of exact genres, instruments, and moods. Our performance in this area validates the effectiveness of our structure-guided data pipeline. Rather than relying on coarse, track-level global tags, our LALM CoT prompting strategy distills rich semantic descriptions tailored specifically for structurally discrete audio segments. Furthermore, by explicitly training the DiT on these localized structured segments, the model coherently learns an exact alignment between local musical transitions and fine-grained text descriptions, leading to highly robust LALM verifiability. 

\textbf{Algorithmic Efficiency over Data Scale:}
While achieving a globally competitive CLAP score of 0.261, the collective strength of S2Accompanist across fidelity and concept-level semantic adherence ultimately secured us the Overall Rank 1 in the Efficiency Track. Our results strongly corroborate the core premise of the ATTM Grand Challenge: with intelligent architectural design and structure-guided data formulation, it is entirely feasible to reduce the reliance on large-scale compute and data. S2Accompanist proves that a lightweight 402M parameter model, trained exclusively on a constrained academic dataset, can achieve highly competitive performance under strict resource constraints, effectively bridging the gap with larger-scale T2M systems.

\subsection{Subjective Evaluation Results}
\label{ssec:sub_eval_results}

Subjective evaluations were conducted by the challenge organizers to further assess the perceptual quality of the generated pure accompaniments. Table \ref{tab:mos_results} presents the Mean Opinion Score (MOS) results on the final test set, evaluated by both general listeners ($\text{MOS}_{\text{all}}$) and expert annotators ($\text{MOS}_{\text{expert}}$). 

\begin{table}[htbp]
\centering
\caption{Subjective Evaluation Results (MOS) released by the ATTM Challenge organizers.}
\label{tab:mos_results}
\resizebox{\columnwidth}{!}{
\begin{tabular}{lccc}
\toprule
\textbf{Model} & \textbf{$\text{MOS}_{\text{all}}$} & \textbf{$\text{MOS}_{\text{expert}}$} & \textbf{Rank} \\ 
\midrule
MusicGen-small \cite{musicgen} & 3.538 & 3.425 & --- \\ 
\midrule
\textbf{S2Accompanist (Ours)} & \textbf{3.250} & \textbf{3.186} & \textbf{Efficiency 1st} \\ 
Submission e01 & 3.225 & 3.177 & Efficiency 2nd \\ 
Submission e08 & 3.119 & 3.044 & Efficiency 3rd \\ 
Submission e05 & 2.969 & 2.929 & --- \\ 
\midrule
Submission p05 & 3.344 & 3.327 & Performance 1st \\ 
Submission p00 & 2.006 & 2.044 & --- \\ 
\bottomrule
\end{tabular}}
\vspace{-4pt}
\end{table}

As shown in Table \ref{tab:mos_results}, S2Accompanist (officially submitted as e07) achieved an $\text{MOS}_{\text{all}}$ of 3.250 and an $\text{MOS}_{\text{expert}}$ of 3.186, ranking first in the Efficiency Track. These subjective scores are consistent with the objective evaluation results presented in Section III-C. This strong subjective performance under strict resource constraints further validates the effectiveness of our proposed methods in generating high-fidelity and musically coherent pure accompaniments.

\section{Ablations}
\label{sec:ablation}

To validate the core components of S2Accompanist, we conduct ablation studies focusing on three key aspects: the generative prior (VAE), the data captioning strategy, and the quality-based fine-tuning stage. It is important to note that these ablation experiments are evaluated on a custom internal test set. This set comprises 200 LLM-generated natural language descriptions designed to closely mirror the style of the official prompts. Consequently, the absolute metric values exhibit slight discrepancies compared to the official benchmark results in Table \ref{tab:main_results}. The quantitative results are summarized in Table \ref{tab:ablation}.

\begin{table}[htbp]
\centering
\caption{Ablation study on VAE architecture, captioning strategy, and training stages. Best results in each block are highlighted.}
\label{tab:ablation}
\resizebox{\columnwidth}{!}{
\begin{tabular}{lccc}
\toprule
\textbf{Model Variant} & \textbf{FAD $\downarrow$} & \textbf{CLAP $\uparrow$} & \textbf{CCS $\uparrow$} \\
\midrule
\multicolumn{4}{l}{\cellcolor[HTML]{E6F2FF}\textit{Effect of Semantic VAE Fine-Tuning (at 100k steps)}} \\
w/ DiffRhythm VAE  & 0.623 & 0.143 & \textbf{0.731} \\
w/ Semantic VAE Fine-Tuning  & \textbf{0.367} & \textbf{0.152} & 0.714 \\
\midrule
\multicolumn{4}{l}{\cellcolor[HTML]{E6F2FF}\textit{Effect of Structure \& Captioning (at 100k steps)}} \\
Track-level Caption & \textbf{0.367} & 0.152 & 0.714 \\
Segment-level Caption & 0.383 & \textbf{0.179} & \textbf{0.793} \\
\midrule
\multicolumn{4}{l}{\cellcolor[HTML]{E6F2FF}\textit{Effect of Quality-Based SFT}} \\
Pretrain Only (400k steps) & 0.348 & 0.182 & 0.745 \\
Pretrain + SFT (5 Epochs) & 0.320 & 0.191 & \textbf{0.805} \\
Pretrain + SFT (10 Epochs) & \textbf{0.301} & \textbf{0.219} & 0.801 \\
\bottomrule
\end{tabular}}
\vspace{-4pt}
\end{table}

\textbf{Effect of Semantic VAE.} We compare our Semantic-Aware VAE against the standard DiffRhythm VAE under identical training configurations (100k steps). As shown in Table \ref{tab:ablation}, integrating the Semantic VAE leads to a notable improvement in audio fidelity, with FAD dropping from 0.623 to 0.367. This confirms that explicitly distilling fundamental musical structures (LeadSheet) into the acoustic latent space is crucial for preserving harmony and reducing generative artifacts in pure music generation.

\textbf{Effect of Structured Captioning.} To verify our structure-guided data formulation, we contrast models trained on global track-level captions versus localized segment-level captions. While the track-level approach provides a slightly lower FAD, shifting to segment-level captioning yields consistent gains in semantic alignment (CLAP improves from 0.152 to 0.179, and CCS surges from 0.714 to 0.793). This demonstrates that localized structural conditions are indispensable for fine-grained musical topic and instrument control.

\textbf{Effect of Quality-Based SFT.} Finally, we evaluate the impact of our dual-metric data grading pipeline. Compared to the pre-training stage (using 100\% data), applying SFT solely on the top 20\% highest-quality segments consistently improves all metrics. At 10 epochs, FAD reaches an optimal 0.301 with strong CLAP (0.219) and CCS (0.801) scores. This indicates that aggressive filtering of separation artifacts and semantic hallucinations during fine-tuning is a highly effective strategy for boosting the upper bound of generation quality.

\section{Conclusion}
\label{sec:conclusion}

In this paper, we present S2Accompanist, an efficient diffusion model tailored for high-fidelity music accompaniment generation. To address the inherent lack of localized annotations in existing pure music datasets, we design a comprehensive automated data pipeline that leverages structural segmentation and LALM-driven fine-grained captioning. Furthermore, to enhance the acoustic latent representation, we propose a Semantic-Aware VAE fine-tuning strategy that distills robust LeadSheet features into the generative prior. Extensive evaluations on the ATTM Grand Challenge benchmark demonstrate that S2Accompanist achieves state-of-the-art objective performance. By securing first place in the Efficiency Track, our approach demonstrates that fine-grained data formulation and semantic-aware structural guidance can effectively reduce dependence on massive computing and large-scale data in text-to-music generation.

\bibliographystyle{IEEEbib}
\bibliography{icme2026references}

\end{document}